\documentclass[twocolumn,aps,prl,showpacs,amssymb,floatfix]{revtex4}
\usepackage{graphicx}
\usepackage{natbib}
\usepackage{times}

\begin{document}

\title{Universal shapes formed by two interacting cracks}

\author{Melissa L. Fender,$^1$ Fr{\'e}d{\'e}ric Lechenault,$^{1,2}$ and Karen E. Daniels$^1$}
\email{kdaniel@ncsu.edu}

\affiliation{$^1$Department of Physics, NC State University, Raleigh, NC, USA, 27695 \\
$^2$LCVN, UMR 5587 CNRS-UM2, Universit\'e Montpellier II, place Eug\`ene Bataillon, 34095 Montpellier, France}

\date{25 May 2010}
\begin{abstract}
We investigate the origins of the widely-observed ``en passant'' crack pattern which forms through interactions between two approaching cracks. A rectangular elastic plate is notched on each long side and then subjected to quasistatic uniaxial strain from the short side. The two cracks propagate along approximately straight paths until they pass each other, after which they curve and release a lenticular fragment. We find that for materials with diverse mechanical properties, the shape of this fragment has an aspect ratio of 2:1, with the length scale set by the initial crack offset $s$ and the time scale set by the ratio of $s$ to the pulling velocity. The cracks have a universal square root shape which we understand using a simple geometric model of the crack-crack interaction.
\end{abstract}

\pacs{62.20.mm, 
46.50.+a, 
}

\maketitle

Brittle failure through multiple cracks occurs in a wide variety of contexts, from microscopic failures in dental enamel \citep{Bechtle-2010-FBD} and cleaved silicon \citep{Gronwald-1984-XTS} to geological faults \citep{Pollard-1984-PLO, Acocella-2000-ILE} and planetary ice crusts \citep{Patterson-2010-SLE}. In each of these situations, with complicated curvature and stress geometries, pairwise interactions between approaching cracks nonetheless produce characteristically-curved fracture paths known in the geologic literature as en passant cracks \citep{Kranz-1979-CCC}. While the fragmentation of solids via many interacting cracks has seen wide investigation \citep{Pollard-1988-PUJ, Allain-1995-RPC, Astrom-1997-FBM,  Astrom-2000-UF, Leung-2000-PFS, Wittel-2004-FS, Jagla-2004-MCP, Bohn-2005-HCP1, Tarafdar-2008-CFD}, less attention has been paid to the details of individual crack-crack interactions. Despite extensive observations of the phenomenon of overlapping cracks in geologic settings \citep{Kranz-1979-CCC, Pollard-1982-FID, Pollard-1984-PLO, Acocella-2000-ILE} as well as controlled laboratory experiments \citep{Swain-1978-OOI, Mills-1980-DGO, Hull-1995-EMM, Cortet-2008-ARC, Acocella-2008-TFO, Tentler-2010-ICO} and simulations \citep{Sempere-1986-OSC, Chan-1991-ECF, Baud-1997-TAP}, an understanding of how similar shapes arise from different dynamics remains lacking.

\begin{figure}
\centerline{\includegraphics[width=\linewidth]{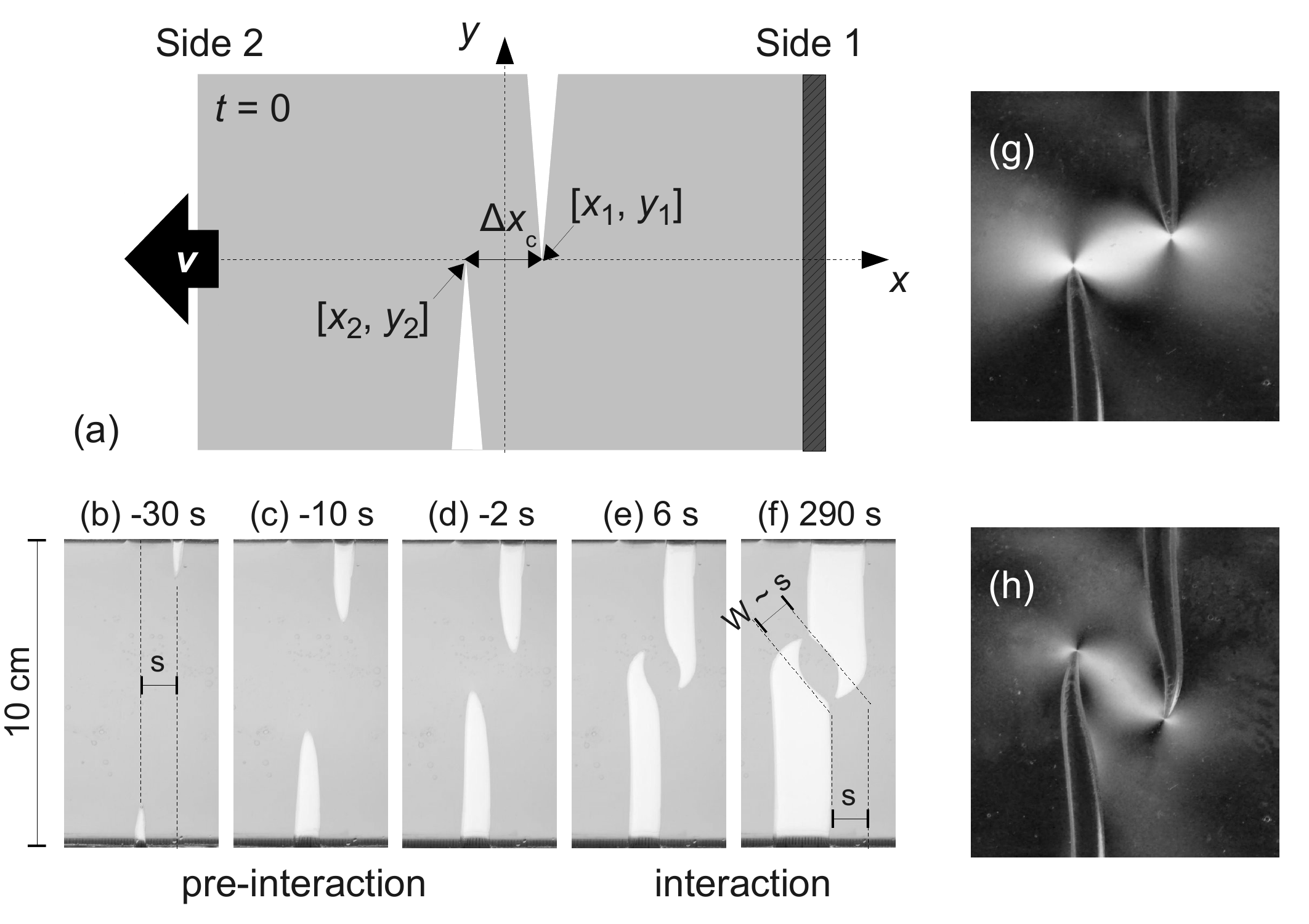}} 
\caption{(a) Experiment schematic, with Side 1 stationary and Side 2 pulled at constant velocity $v$. The point at which the two crack paths first cross their mutual perpendicular sets $t = 0$, and $[x,y] = 0$. (b--f) Detail of gelatin sheet showing dynamics. Detail of photoelastic response of gelatin sheet (g) before and (h) during crack curvature.}
\label{setup} 
\end{figure}

Through experiments and a simple geometric model, we elucidate the mechanism behind these two-crack interactions and quantify the scaling properties which allow for the same shape in diverse situations. We choose a  geometry for which two-crack interactions are the dominant failure mode of the material: a uniaxial strain test on two prepared notches on opposite lateral sides of an elastic sample. As the ends of the sample are pulled apart quasistatically, the two notches propagate inward as single cracks until they pass and begin to curve towards each other, cutting out a lenticular fragment (see Fig.~\ref{setup}b-f). The direction of this curvature is controlled by the ratio of mode I and mode II stress intensity factors \citep{Pollard-1982-FID}. 

Using materials varying in bulk modulus, Poisson ratio, and heterogeneity, we uncover a universal, scale-invariant shape and dynamics for this two-crack interaction. The shape of the final fragment exhibits rate-, material- and scale-invariance, with an aspect ratio (length:width) of $\Gamma = 2$ and an approximately square root shape which begins at the point where the two cracks pass each other during propagation. We observe that the length scale of the fracture curvature is set by only the initial notch offset $s$, and that the time scale is set by the ratio of $s$ to the pulling velocity $v$. In order to explain this universal shape, we construct a simple geometric model for the interaction dynamics based on the observed stress axes. This simple model provides a quantitative prediction which reproduces both the shape of the lens and its aspect ratio.

\paragraph{Experiment:}
We perform experiments on sheets of various materials in a motorized frame which pulls one edge of the sheet away from the opposite edge. To start, we cut a pair of offset notches separated by a distance $s$; these two cuts instigate the cracks which propagate inward and interact under uniaxial strain until they cut out a lenticular fragment, visible in Fig.~\ref{setup}f. The sheets are approximately $10 \times 20$~cm$^2$ and pulling velocities are 0.7 to 14 mm/s. We collect data on the crack propagation and interaction using three techniques: shape analysis of the final fragment, crack-tip tracking to capture the dynamics, and photoelastic imaging of the stress axes. For the final shape analysis, we remove the fragment from the apparatus and perform measurements on an unstressed sample. We track the location of the tip on a sequence of images taken every $4$~s 
(see Fig.~\ref{setup}b-f); during this tracking, the sample remains in a stressed state. For the gelatin samples, we can use a polariscope to visualize the internal stresses during fracture \citep{Simha-1986-DPI}, as shown in Fig.~\ref{setup}gh.

The experiments investigate the effect of the initial crack offset $s$, thickness $h$, material properties, and pulling velocity $v$. Our primary material is
$7\%$w gelatin ($E \sim10^5$~Pa, $\nu = 0.4$ to $0.5$, $h=8$~mm) \citep{Markidou-2005-SME}, 
	with comparison runs in 
nitrile ($E \sim 10^6$~Pa, $\nu \sim 0.5$, $h=0.1$~mm) \citep{Kakavas-1996-MPB},
cork ($E \sim 10^7$~Pa, $\nu \sim 0$, $h=3$ or $6$~mm),
polystyrene foam ($E \sim 10^9$~Pa, $\nu \sim 0$, $h=6$~mm) \citep{Rinde-1970-PRR},
and aluminum foil ($E \sim 10^{11}$~Pa, $\nu \sim 0.3$, $h=0.02$~mm).
These materials are selected for their diversity of material properties, and cover the full range of Poisson ratio values from $0$ to $0.5$. For the gelatin, which is weak enough to break under its own weight, we support the sheet from below using a clear sheet lubricated with vegetable oil or water to minimize the frictional interaction with the substrate. The ends of the gelatin sheets are affixed to the apparatus by means of mesh brackets cast into the sheet during preparation. All other materials are clamped at their ends and hang under their own weight; as a result, the sheet is not constrained to remain in a single plane during fracture.

\begin{figure}
\centerline{\includegraphics[width=\linewidth]{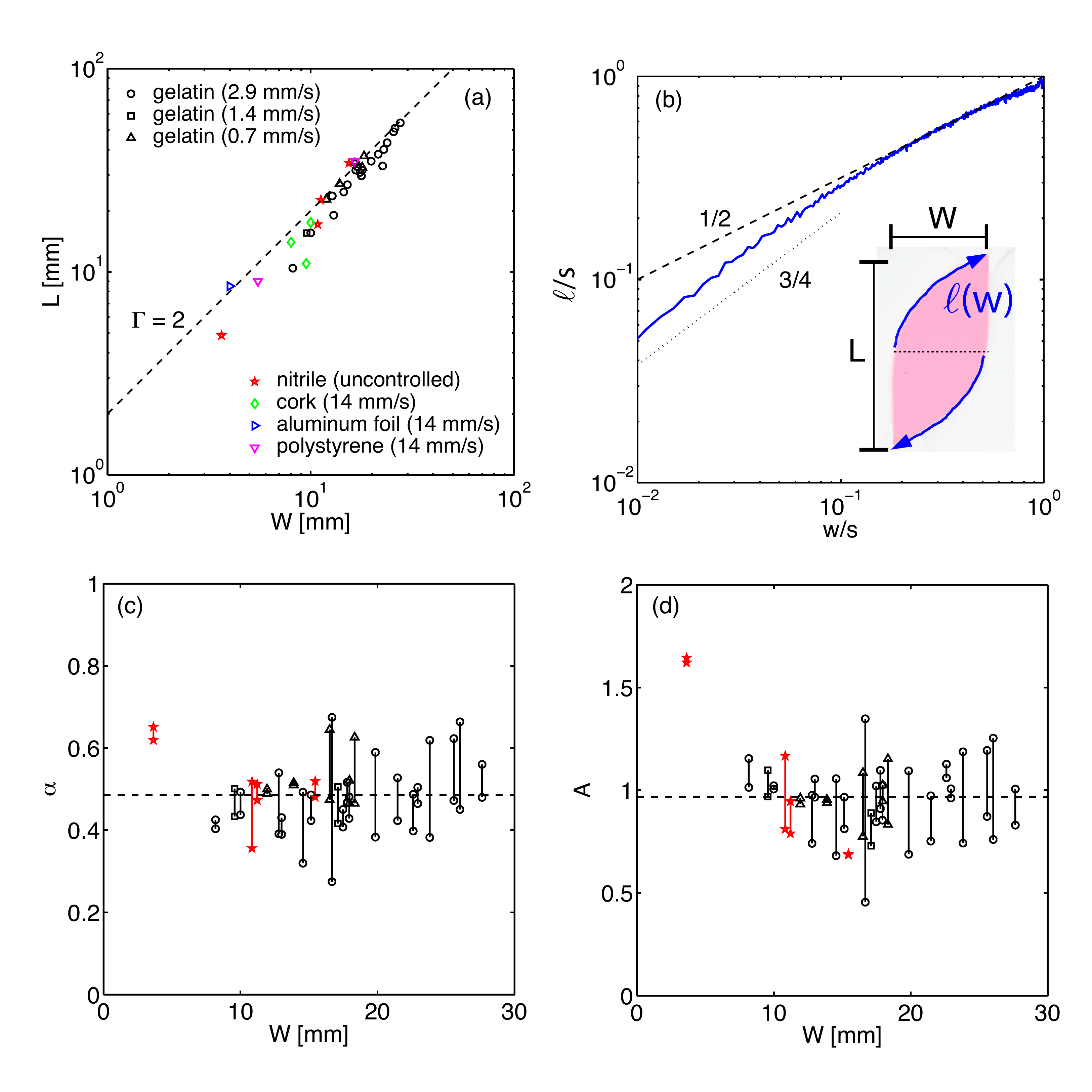}}
\caption{[Color Online] (a) Measurement of aspect ratio $\Gamma$ of final fragment shapes for constant-$v$ fracture of gelatin, cork, polystyrene foam, and aluminum foil, and uncontrolled fracture of nitrile. Dotted line is $L = 2 W$. (b) Average of edges $\ell(w)$ for 8 gelatin fragments, with both lengths scaled by $s$. Dashed line is Eq.~\ref{e:powerlaw} with $A=1$ and  $\alpha=\frac{1}{2}$, which corresponds to the model (Eq.~\ref{e:model}). Inset: Sample image of fragment showing length $L$, width $W \approx s$, and extracted edge $\ell(w)$. (c) Power law exponent $\alpha$ as a function of $W$, obtained from fitting  Eq.~\ref{e:powerlaw}; dashed line is mean $\langle \alpha \rangle = 0.49$. (d)  Power law prefactor $A$ as a function of $W$, obtained from fitting Eq.~\ref{e:powerlaw} with fixed $\alpha = \frac{1}{2}$; dashed line is mean $\langle A \rangle  = 0.97$. In (c,d) the paired points connected by vertical bars correspond to the two sides of a single fragment. All fits to Eq.~\ref{e:powerlaw} are performed over the portion of $\ell(w)$ with $w/s > 0.1$ }
\label{shape} 
\end{figure}

\paragraph{Shape:} 
For each final fragment, we measure the length $L$ and width $W$, with $W \approx s$ corresponding to the initially-parallel crack edges (see Fig.~\ref{setup}f). As shown in Fig.~\ref{shape}a, we observe an aspect ratio $\Gamma \equiv L/W \approx 2$, independent of $s$, material, and $v$. Remarkably, $\Gamma$ is constant even for runs in which the sheet curved out of plane, started from non-parallel initial cracks, or was subjected to shear in addition to the tensile strain. For the last two cases, the cracks reoriented to two straight paths perpendicular to the direction of loading, thereby still resulting a fragment with $\Gamma \approx 2$.

To examine the universality of the shape which resulted in $\Gamma = 2$, we examine 8 gelatin fragments scanned at 600 dpi resolution. For each, we extract the coordinates $[\ell, w]$  along the edge connecting the midpoint of the fragment to the tip, as shown in Fig.~\ref{shape}b. Note that the two initial cracks (separated by $s$) remain approximately parallel until they reach this midpoint, after which they curve towards each other along the path $\ell(w)$. These dynamics will be further quantified below. We examine the shape of the fragment by scaling both coordinates of $\ell(w)$ by $s$ and averaging over the ensemble of fragments. Each half-fragment spans $0 < w < 1$ and $0 < \ell < 1$; the full length is $2\ell$, corresponding to $\Gamma = 2$. The average shape $\langle \ell(w) \rangle$ has an approximately power-law shape over much of its length, with the exponent varying from $\frac{3}{4}$ initially to $\frac{1}{2}$ for $w/s \gtrsim 0.1$. 

Motivated by this average shape, we fit each edge (two from each fragment) to the functional form
\begin{equation}
\left( \frac{\ell}{s} \right)=A \left( \frac{w}{s} \right)^{\alpha}
\label{e:powerlaw}
\end{equation}
for the region $w/s > 0.1$ and find mean values of $\langle A \rangle = 0.97 \pm 0.03$ and $\langle \alpha \rangle = 0.49 \pm 0.01$, with no systematic dependence on $s$, material properties, or $v$ (see Fig.~\ref{shape}cd). Individual fragments vary from this mean behavior, particularly when the initial cracks were imperfectly parallel or the gelatin sheet was poorly lubricated. Thus, we conclude that the initial crack offset $s$ sets not only $L$ and $W$ of the final fragment, but is the only input to the universal shape function given in Eq.~\ref{e:powerlaw}.

\paragraph{Dynamics:}
To characterize the dynamics which result in this characteristic shape, we track the two crack tip positions for runs started from different $s$ and pulled at three different $v$. We record the coordinates  $[x_{i}(t),y_{i}(t)]$, where $i=1$ for the stationary side and $i=2$ for the pulled side. We set the origin of our moving coordinate system so that $[x, y] = 0$ at the midpoint between the two tips, and take $t=0$ at the interpolated time when the two crack paths cross $y=0$. The offset between the crack tips is measured at $t=0$ to be $\Delta x_c \equiv x_1(0) - x_2(0) \approx s$, which can be influenced by slight repulsion near $t=0$ \citep{Pollard-1982-FID, Cortet-2008-ARC}. In addition, the characteristic timescale $\tau \equiv \Delta x_c / v$  is used to make comparisons between runs at different $v$, and we scale all distances by $\Delta x_c$,  The resulting scaled variables  are
\begin{equation}
\tilde{t} = \frac {t}{\tau}, \hspace{8mm}
\tilde{x_i} = \frac{x_i}{\Delta x_c}, \hspace{8mm}
\tilde{y_i}  = \frac{y_i}{\Delta x_c}.  
\label{e:scaling}
\end{equation}
In tracking the dynamics of the tip, all measurements are made on a stressed sample.

\begin{figure}
\includegraphics[width=\linewidth]{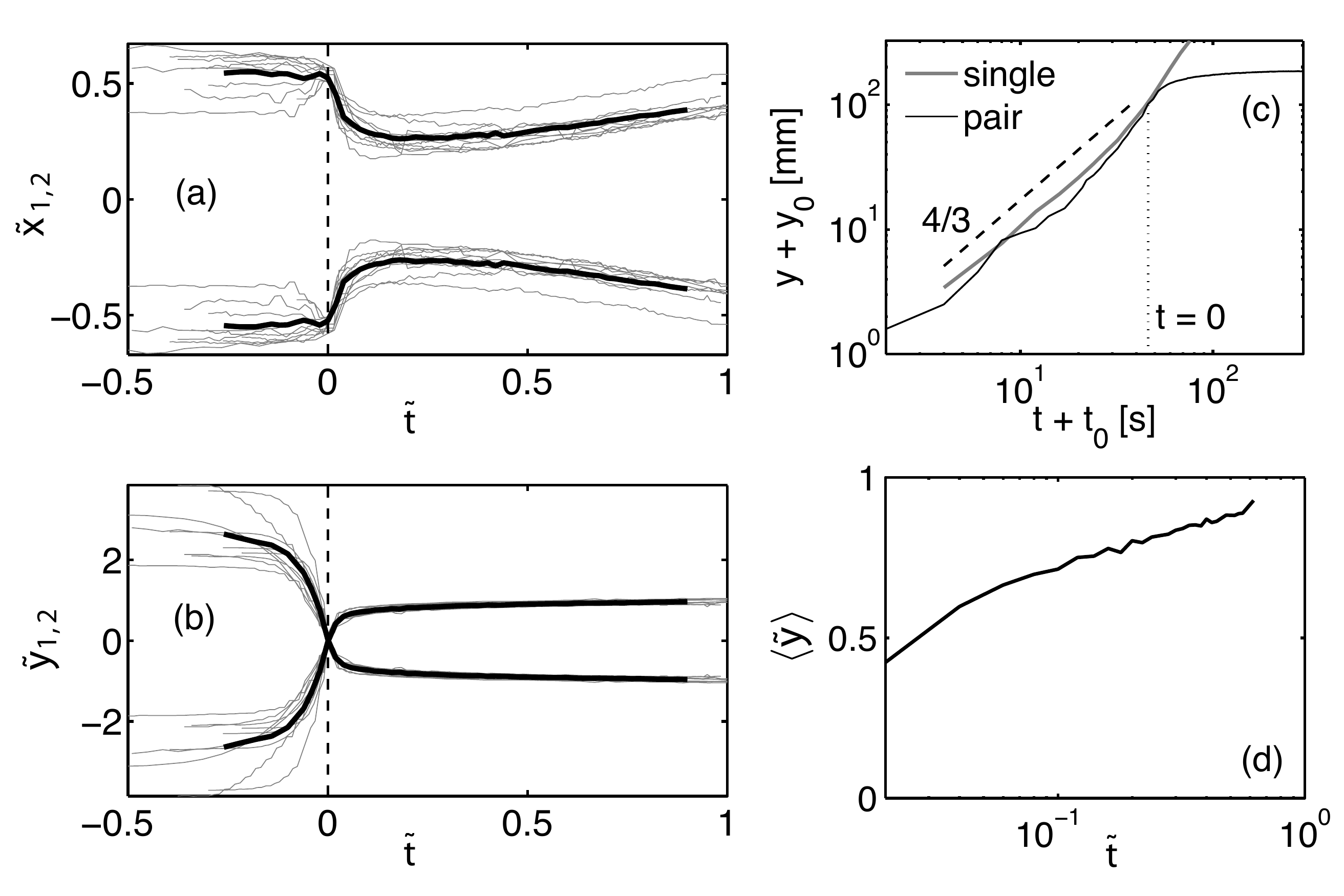}
\caption{(a) ${\tilde x}_i(\tilde t)$ and (b) ${\tilde y}_i(\tilde t)$  for 12 experiments on gelatin at $v = 0.7$, 1.4 and 2.9 mm/s. (c) Pre-interaction stage: $y_1(t)$ for a single-crack ($v=0.7$~mm/s) and a double-crack  ($v=1.4$~mm/s). $y_0<0$ is the starting coordinate of each crack at time $t_0<0$. (d) Interaction stage:  $\langle \tilde y(\tilde t) \rangle$.}
\label{traj} 
\end{figure}

Fig.~\ref{traj}ab examines both individual (thin lines) and average (thick lines) trajectories, demonstrating that the scaling in Eq.~\ref{e:scaling} collapses the curves into a universal shape; no systematic dependence on $\Delta x_c$ or $v$ is present in the scatter. Note that ${\tilde x}(\tilde t)$ is on average constant for $\tilde t<0$, indicating that little crack curvature is taking place at this stage. This results in each final fragment containing two parallel edges at opposite ends of the fragment.  In addition, we observe that each crack initially behaves as if it were a single, isolated crack pulled at half the velocity, as shown in Fig.~\ref{traj}c. For either a single crack or one of a crack pair, $y\propto (t + t_0)^{4/3}$. Therefore, we refer to $\tilde t<0$ as the pre-interaction stage and $\tilde t>0$ as the interaction stage.  

To examine how the observed similarity in $\ell(w)$ arises from the dynamics, we consider $\langle {\tilde x}_1(\tilde t) \rangle $ and $\langle {\tilde y}_1(\tilde t) \rangle$, where $\langle \cdot \rangle$ is the average over the 12 examples.  During the interaction stage, $\langle \tilde y \rangle$ is observed to grow only logarithmically (see Fig.~\ref{traj}d). This slowing down can be understood in terms of the rotation of the fragment, rather than fracture, becoming the dominant means of accommodating strain the thinner the remaining connection becomes (see Fig.~\ref{setup}f.)

These dynamics highlight several important features of the en passant crack geometry which may be quite general. First, the interaction between the cracks starts only when the tips pass each other. From that point, all length scales are set by the crack offset $s$ and all time scales are set by the ratio of this offset to $v$. The final shape has an aspect ratio $\Gamma = 2$ for materials with diverse elastic properties, including elastic modulus, Poisson ratio and heterogeneity. The curved portion of the fragment takes a power law shape with an exponent close to $\frac{1}{2}$.

This universal shape suggests that a geometric model can provide insight into the dynamics, without reference to the full elastic problem. Initially, the crack paths remain straight and parallel because there is a large bulk region between the cracks, allowing the principal axis of the stress to be aligned with the boundary loading. This is pure mode I fracture, and proceeds identically to a single crack, without torque (see Fig.~\ref{traj}c). However, once the crack tips pass each other, the principal axis of the stress within the central region rotates and connects the tips. This provides a net torque on the central part of the sample and leads to locally mixed mode I and II loading. At each strain increment, the line connecting the two tips rotates and the subsequent fracture occurs relative to this new line. These features can be observed in the polariscope images of the internal stresses show in Fig.~\ref{setup}gh.

\paragraph{Model:} 
We construct a simple geometric model of the two-crack interaction for $\tilde t >0$ starting from three assumptions. First, that the trace of the stress tensor is is largest along the line connecting the two cracks. Second, that each crack propagates orthogonal to this direction. This corresponds to the principle of local symmetry, whereby mode I is selected over mode II \citep{Goldstein-1974-BFS, Audoly-2005-CTS}. These two assumptions alone are sufficient to generate curvature in the crack tip trajectory: once the cracks have moved past each other, the line connecting the tips rotates, and the crack direction rotates with it. For simplicity, we also assume that the two tips remain laterally separated by their initial offset $s$. As can be seen from Fig.~\ref{traj}a, $\tilde x$ remains in the vicinity of $\frac{1}{2}$ for ${\tilde t} > 0$, but with additional dynamics which we do not capture with the simple model.

\begin{figure}
\includegraphics[width=\linewidth]{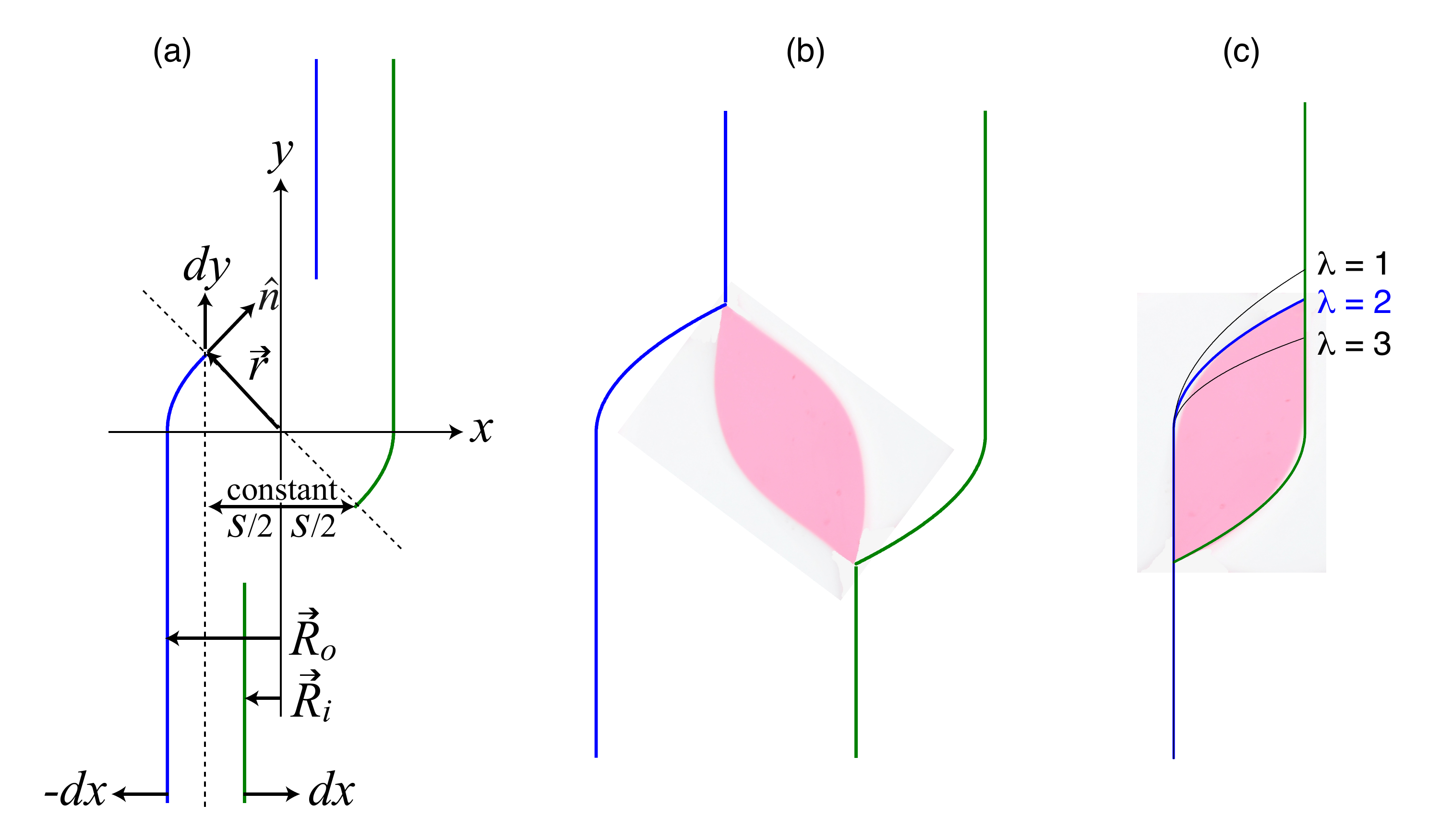}
\caption{[Color Online] (a) Model solution (thick lines) during interaction stage (${\tilde t}>0$), with definitions of variables. Model in (b) final state and (c) translated from final state to show crack edges superimposed on experimental image. Comparison to $\lambda=1,3$ shown as thin lines in (c).}
\label{model} 
\end{figure}

To formulate this model algebraically, we establish a coordinate system (see Fig.~\ref{model}a) around the center of symmetry. Within this frame, the two edges are pulled away at equal and opposite velocities. During each strain interval, each outer crack edge (${\vec R}_o$) moves a distance $dx$ away from the center. By continuity, each inner edge ($\vec R_i$) moves in the opposite direction by the same amount. Each crack propagates forward to relieve the resulting stress, with the direction set by the orthogonality condition and a distance chosen to keep each tip at fixed $x-$coordinate $\pm\frac{s}{2}$. At each strain interval, the coordinates of the left-side tip are $\vec r  \equiv [-\frac{s}{2}, y]$ and the cracking direction is normal to this vector, in the direction ${\vec n} = [1, \frac{s}{2y}]$. Thus, for each outward displacement by $dx$, the outer edge  $\vec R_o \equiv [x, 0]$  will move leftward by $-dx$ so that  $x \rightarrow x - dx$, the inner edge  $\vec R_i \equiv [-(s+x), 0]$ will move rightward by $dx$, and the $y$ coordinate of the tip will advance by $y \rightarrow y + \frac{sdx}{2y}$ to correspondingly relieve the strain. This provides the differential equation 
\begin{equation}
-\frac{dy}{dx}=\frac{s}{2y}.
\label{ODE}
\end{equation}
Integration yields $(\frac{y}{s})^2 = -\left(\frac{1}{2}+\frac{x}{s}\right)$, where the constant of integration is set by $y(-\frac{s}{2}) = 0$. To compare with the experimental results, we define $\tilde{y} \equiv \frac{y}{s}$ and $\tilde{x} \equiv -(\frac{1}{2}+\frac{x}{s})$, which leads to the simple shape
\begin{equation}
\tilde{y} = \tilde{x}^{\frac{1}{2}}.
\label{e:model}
\end{equation}
The crack propagation stops when the inner crack edge meets the tip of the other crack, which sits at $\frac{s}{2}$. This occurs when $-(s+x_f)=\frac{s}{2}$, or equivalently $\tilde{x}_f = \tilde{y}_f = 1$. Importantly, this solution matches the $\Gamma=2$ shown in Fig.~\ref{shape}a and $\langle A \rangle = 0.97$ and $\langle \alpha \rangle = 0.49$ shown in Fig.~\ref{shape}cd. For comparison, we have superimposed the model prediction on an experimentally-obtained image in Fig.~\ref{model}c and find excellent agreement except at the early stages of curvature. This early regime corresponds to the portion of the shape with $\alpha=\frac{3}{4}$ in Fig.~\ref{shape}b, and to the time immediately after $\tilde t =0$  Fig.~\ref{traj}a.  

While we have constructed the model in terms of boundary-driven strain increments $dx$, Eq.~\ref{ODE} and the same final shape would result for any brittle fracture mechanism, including cyclic thermal expansion and contraction, as long as the crack growth still occurred in the direction normal to the line separating two tips which remained at fixed separation.
For a boundary loading which provides a different stress intensity factor ratio than observed here, the crack may grow at a larger or smaller (non-orthogonal) angle with respect to $\vec r$ \citep{Pollard-1982-FID}. In such a case, the factor $2$ in  Eq.~\ref{ODE} can be generalized to a constant $\lambda$ with $-\frac{dy}{dx}=\frac{s}{\lambda y}$. When integrated, the solutions for $\lambda \ne 2$ retain $\alpha = \frac{1}{2}$ but have $A = A(\lambda)$ and therefore $\Gamma \ne 2$, as shown in Fig.~\ref{model}c. This last case is noteworthy, since in geologic observations, $\Gamma > 2$ is commonly observed \citep{Acocella-2000-ILE}. In experiments where we provide additional lateral boundary stresses, we are able to achieve both larger and smaller aspect ratios. This raises the possibility of using lenticular crack shapes as a diagnostic for the stress conditions under which cracks were formed in nature \citep{Olson-1989-IPN}. In particular, $\Gamma$ may serve as a means to infer the boundary loading in situations where history and dynamics are inaccessible, for instance in observations of Europa's ice sheets \citep{Patterson-2010-SLE}.

In conclusion, we have identified the length and time scales which control a generic two-crack interaction problem. A simple geometric model allows us to identify the key mechanism leading to the universal shape: the largest tensile stress follows a straight line connecting the tips and the cracks propagate perpendicular to this line.

\paragraph{Acknowledgments:} The investigators are grateful to Pedro Reis and Beno{\^i}t Roman for instigating this investigation, and for later contributions regarding interpretation, and to Matteo Ciccotti, Nicholas Hayman, James Puckett and Michael Shearer for useful discussions. We thank NSF for support under grants DMS-0604047 and DMR-0644743.



\begin{thebibliography}{32}
\expandafter\ifx\csname natexlab\endcsname\relax\def\natexlab#1{#1}\fi
\expandafter\ifx\csname bibnamefont\endcsname\relax
  \def\bibnamefont#1{#1}\fi
\expandafter\ifx\csname bibfnamefont\endcsname\relax
  \def\bibfnamefont#1{#1}\fi
\expandafter\ifx\csname citenamefont\endcsname\relax
  \def\citenamefont#1{#1}\fi
\expandafter\ifx\csname url\endcsname\relax
  \def\url#1{\texttt{#1}}\fi
\expandafter\ifx\csname urlprefix\endcsname\relax\def\urlprefix{URL }\fi
\providecommand{\bibinfo}[2]{#2}
\providecommand{\eprint}[2][]{\url{#2}}

\bibitem[{\citenamefont{Bechtle et~al.}(2010)\citenamefont{Bechtle, Habelitz,
  Klocke, Fett, and Schneider}}]{Bechtle-2010-FBD}
\bibinfo{author}{\bibfnamefont{S.}~\bibnamefont{Bechtle}},
  \bibinfo{author}{\bibfnamefont{S.}~\bibnamefont{Habelitz}},
  \bibinfo{author}{\bibfnamefont{A.}~\bibnamefont{Klocke}},
  \bibinfo{author}{\bibfnamefont{T.}~\bibnamefont{Fett}}, \bibnamefont{and}
  \bibinfo{author}{\bibfnamefont{G.~A.} \bibnamefont{Schneider}},
  \bibinfo{journal}{Biomaterials} \textbf{\bibinfo{volume}{31}},
  \bibinfo{pages}{375} (\bibinfo{year}{2010}).

\bibitem[{\citenamefont{Gronwald and Henzler}(1984)}]{Gronwald-1984-XTS}
\bibinfo{author}{\bibfnamefont{K.~D.} \bibnamefont{Gronwald}} \bibnamefont{and}
  \bibinfo{author}{\bibfnamefont{M.}~\bibnamefont{Henzler}},
  \bibinfo{journal}{Appl. Phys. A} \textbf{\bibinfo{volume}{34}},
  \bibinfo{pages}{253} (\bibinfo{year}{1984}).

\bibitem[{\citenamefont{Pollard and Aydin}(1984)}]{Pollard-1984-PLO}
\bibinfo{author}{\bibfnamefont{D.~D.} \bibnamefont{Pollard}} \bibnamefont{and}
  \bibinfo{author}{\bibfnamefont{A.}~\bibnamefont{Aydin}},
  \bibinfo{journal}{J. Geophys. Res.}
  \textbf{\bibinfo{volume}{89}}, \bibinfo{pages}{17} (\bibinfo{year}{1984}).

\bibitem[{\citenamefont{Acocella et~al.}(2000)\citenamefont{Acocella,
  Gudmundsson, and Funiciello}}]{Acocella-2000-ILE}
\bibinfo{author}{\bibfnamefont{V.}~\bibnamefont{Acocella}},
  \bibinfo{author}{\bibfnamefont{A.}~\bibnamefont{Gudmundsson}},
  \bibnamefont{and}
  \bibinfo{author}{\bibfnamefont{R.}~\bibnamefont{Funiciello}},
  \bibinfo{journal}{J. Struct. Geol.}
  \textbf{\bibinfo{volume}{22}}, \bibinfo{pages}{1233} (\bibinfo{year}{2000}).

\bibitem[{\citenamefont{Patterson and Head}(2010)}]{Patterson-2010-SLE}
\bibinfo{author}{\bibfnamefont{G.~W.} \bibnamefont{Patterson}}
  \bibnamefont{and} \bibinfo{author}{\bibfnamefont{J.~W.} \bibnamefont{Head}},
  \bibinfo{journal}{Icarus} \textbf{\bibinfo{volume}{205}},
  \bibinfo{pages}{528} (\bibinfo{year}{2010}).

\bibitem[{\citenamefont{Kranz}(1979)}]{Kranz-1979-CCC}
\bibinfo{author}{\bibfnamefont{R.~L.} \bibnamefont{Kranz}},
  \bibinfo{journal}{Int. J. Rock Mech. Min. Sci. \& Geomech. Abstr.}
  \textbf{\bibinfo{volume}{16}}, \bibinfo{pages}{37} (\bibinfo{year}{1979}).

\bibitem[{\citenamefont{Pollard and Aydin}(1988)}]{Pollard-1988-PUJ}
\bibinfo{author}{\bibfnamefont{D.~D.} \bibnamefont{Pollard}} \bibnamefont{and}
  \bibinfo{author}{\bibfnamefont{A.}~\bibnamefont{Aydin}},
  \bibinfo{journal}{Geol. Soc. Am. Bull.}
  \textbf{\bibinfo{volume}{100}}, \bibinfo{pages}{1181} (\bibinfo{year}{1988}).

\bibitem[{\citenamefont{Allain and Limat}(1995)}]{Allain-1995-RPC}
\bibinfo{author}{\bibfnamefont{C.}~\bibnamefont{Allain}} \bibnamefont{and}
  \bibinfo{author}{\bibfnamefont{L.}~\bibnamefont{Limat}},
  \bibinfo{journal}{Phys. Rev. Lett.} \textbf{\bibinfo{volume}{74}},
  \bibinfo{pages}{2981} (\bibinfo{year}{1995}).

\bibitem[{\citenamefont{Astrom and Timonen}(1997)}]{Astrom-1997-FBM}
\bibinfo{author}{\bibfnamefont{J.}~\bibnamefont{Astrom}} \bibnamefont{and}
  \bibinfo{author}{\bibfnamefont{J.}~\bibnamefont{Timonen}},
  \bibinfo{journal}{Phys. Rev. Lett.} \textbf{\bibinfo{volume}{79}},
  \bibinfo{pages}{3684} (\bibinfo{year}{1997}).

\bibitem[{\citenamefont{Astrom et~al.}(2000)\citenamefont{Astrom, Holian, and
  Timonen}}]{Astrom-2000-UF}
\bibinfo{author}{\bibfnamefont{J.~A.} \bibnamefont{Astrom}},
  \bibinfo{author}{\bibfnamefont{B.~L.} \bibnamefont{Holian}},
  \bibnamefont{and} \bibinfo{author}{\bibfnamefont{J.}~\bibnamefont{Timonen}},
  \bibinfo{journal}{Phys. Rev. Lett.} \textbf{\bibinfo{volume}{84}},
  \bibinfo{pages}{3061} (\bibinfo{year}{2000}).

\bibitem[{\citenamefont{Leung and Neda}(2000)}]{Leung-2000-PFS}
\bibinfo{author}{\bibfnamefont{K.~T.}~\bibnamefont{Leung}} \bibnamefont{and}
  \bibinfo{author}{\bibfnamefont{Z.}~\bibnamefont{Neda}},
  \bibinfo{journal}{Phys. Rev. Lett.} \textbf{\bibinfo{volume}{85}},
  \bibinfo{pages}{662} (\bibinfo{year}{2000}).

\bibitem[{\citenamefont{Wittel et~al.}(2004)\citenamefont{Wittel, Kun,
  Herrmann, and Kroplin}}]{Wittel-2004-FS}
\bibinfo{author}{\bibfnamefont{F.}~\bibnamefont{Wittel}},
  \bibinfo{author}{\bibfnamefont{F.}~\bibnamefont{Kun}},
  \bibinfo{author}{\bibfnamefont{H.~J.} \bibnamefont{Herrmann}},
  \bibnamefont{and} \bibinfo{author}{\bibfnamefont{B.~H.}
  \bibnamefont{Kroplin}}, \bibinfo{journal}{Phys. Rev. Lett.}
  \textbf{\bibinfo{volume}{93}}, \bibinfo{pages}{035504}
  (\bibinfo{year}{2004}).

\bibitem[{\citenamefont{Jagla}(2004)}]{Jagla-2004-MCP}
\bibinfo{author}{\bibfnamefont{E.~A.} \bibnamefont{Jagla}},
  \bibinfo{journal}{Phys. Rev. E} \textbf{\bibinfo{volume}{69}},
  \bibinfo{pages}{056212} (\bibinfo{year}{2004}).

\bibitem[{\citenamefont{Bohn et~al.}(2005)\citenamefont{Bohn, Pauchard, and
  Couder}}]{Bohn-2005-HCP1}
\bibinfo{author}{\bibfnamefont{S.}~\bibnamefont{Bohn}},
  \bibinfo{author}{\bibfnamefont{L.}~\bibnamefont{Pauchard}}, \bibnamefont{and}
  \bibinfo{author}{\bibfnamefont{Y.}~\bibnamefont{Couder}},
  \bibinfo{journal}{Phys. Rev. E} \textbf{\bibinfo{volume}{71}},
  \bibinfo{pages}{046214} (\bibinfo{year}{2005}).

\bibitem[{\citenamefont{Tarafdar and Sinha}(2008)}]{Tarafdar-2008-CFD}
\bibinfo{author}{\bibfnamefont{S.}~\bibnamefont{Tarafdar}} \bibnamefont{and}
  \bibinfo{author}{\bibfnamefont{S.}~\bibnamefont{Sinha}},
  \bibinfo{journal}{Ind. Eng. Chem. Res.}
  \textbf{\bibinfo{volume}{47}}, \bibinfo{pages}{6459} (\bibinfo{year}{2008}).

\bibitem[{\citenamefont{Pollard et~al.}(1982)\citenamefont{Pollard, Segall, and
  Delaney}}]{Pollard-1982-FID}
\bibinfo{author}{\bibfnamefont{D.~D.} \bibnamefont{Pollard}},
  \bibinfo{author}{\bibfnamefont{P.}~\bibnamefont{Segall}}, \bibnamefont{and}
  \bibinfo{author}{\bibfnamefont{P.~T.} \bibnamefont{Delaney}},
  \bibinfo{journal}{Geol. Soc. Am. Bull. }
  \textbf{\bibinfo{volume}{93}}, \bibinfo{pages}{1291} (\bibinfo{year}{1982}).

\bibitem[{\citenamefont{Swain and Hagan}(1978)}]{Swain-1978-OOI}
\bibinfo{author}{\bibfnamefont{M.~V.} \bibnamefont{Swain}} \bibnamefont{and}
  \bibinfo{author}{\bibfnamefont{J.~T.} \bibnamefont{Hagan}},
  \bibinfo{journal}{ Eng. Fract. Mech.}
  \textbf{\bibinfo{volume}{10}}, \bibinfo{pages}{299} (\bibinfo{year}{1978}).

\bibitem[{\citenamefont{Mills and Walker}(1980)}]{Mills-1980-DGO}
\bibinfo{author}{\bibfnamefont{N.~J.} \bibnamefont{Mills}} \bibnamefont{and}
  \bibinfo{author}{\bibfnamefont{N.}~\bibnamefont{Walker}},
  \bibinfo{journal}{Eng. Fract. Mech.}
  \textbf{\bibinfo{volume}{13}}, \bibinfo{pages}{479} (\bibinfo{year}{1980}).

\bibitem[{\citenamefont{Hull}(1995)}]{Hull-1995-EMM}
\bibinfo{author}{\bibfnamefont{D.}~\bibnamefont{Hull}}, \bibinfo{journal}{Int. J.
  Fract.} \textbf{\bibinfo{volume}{70}}, \bibinfo{pages}{59}
  (\bibinfo{year}{1995}).

\bibitem[{\citenamefont{Cortet et~al.}(2008)\citenamefont{Cortet, Huillard,
  Vanel, and Ciliberto}}]{Cortet-2008-ARC}
\bibinfo{author}{\bibfnamefont{P.-P.} \bibnamefont{Cortet}},
  \bibinfo{author}{\bibfnamefont{G.}~\bibnamefont{Huillard}},
  \bibinfo{author}{\bibfnamefont{L.}~\bibnamefont{Vanel}}, \bibnamefont{and}
  \bibinfo{author}{\bibfnamefont{S.}~\bibnamefont{Ciliberto}},
  \bibinfo{journal}{J. Stat. Mech.}
  \bibinfo{pages}{P10022} (\bibinfo{year}{2008}).

\bibitem[{\citenamefont{Acocella}(2008)}]{Acocella-2008-TFO}
\bibinfo{author}{\bibfnamefont{V.}~\bibnamefont{Acocella}},
  \bibinfo{journal}{ Earth Planet. Sci. Lett.  }
  \textbf{\bibinfo{volume}{265}}, \bibinfo{pages}{379} (\bibinfo{year}{2008}).

\bibitem[{\citenamefont{Tentler and Acocella}(2010)}]{Tentler-2010-ICO}
\bibinfo{author}{\bibfnamefont{T.}~\bibnamefont{Tentler}} \bibnamefont{and}
  \bibinfo{author}{\bibfnamefont{V.}~\bibnamefont{Acocella}},
  \bibinfo{journal}{J. Geophys. Res.}
  \textbf{\bibinfo{volume}{115}}, \bibinfo{pages}{B01401}
  (\bibinfo{year}{2010}).

\bibitem[{\citenamefont{Sempere and Macdonald}(1986)}]{Sempere-1986-OSC}
\bibinfo{author}{\bibfnamefont{J.~C.} \bibnamefont{Sempere}} \bibnamefont{and}
  \bibinfo{author}{\bibfnamefont{K.~C.} \bibnamefont{Macdonald}},
  \bibinfo{journal}{Tectonics} \textbf{\bibinfo{volume}{5}},
  \bibinfo{pages}{151} (\bibinfo{year}{1986}).

\bibitem[{\citenamefont{Chan}(1991)}]{Chan-1991-ECF}
\bibinfo{author}{\bibfnamefont{H.~C.~M.} \bibnamefont{Chan}},
  \bibinfo{journal}{Eng. Fract. Mech.}
  \textbf{\bibinfo{volume}{39}}, \bibinfo{pages}{433} (\bibinfo{year}{1991}).

\bibitem[{\citenamefont{Baud and Reuschle}(1997)}]{Baud-1997-TAP}
\bibinfo{author}{\bibfnamefont{P.}~\bibnamefont{Baud}} \bibnamefont{and}
  \bibinfo{author}{\bibfnamefont{T.}~\bibnamefont{Reuschle}},
  \bibinfo{journal}{Geophys. J. Int. }
  \textbf{\bibinfo{volume}{130}}, \bibinfo{pages}{460} (\bibinfo{year}{1997}).

\bibitem[{\citenamefont{Simha et~al.}(1986)\citenamefont{Simha, Fourney,
  Barker, and Dick}}]{Simha-1986-DPI}
\bibinfo{author}{\bibfnamefont{K.~R.~Y.} \bibnamefont{Simha}},
  \bibinfo{author}{\bibfnamefont{W.~L.} \bibnamefont{Fourney}},
  \bibinfo{author}{\bibfnamefont{D.~B.} \bibnamefont{Barker}},
  \bibnamefont{and} \bibinfo{author}{\bibfnamefont{R.~D.} \bibnamefont{Dick}},
  \bibinfo{journal}{Eng. Fract. Mech.}
  \textbf{\bibinfo{volume}{23}}, \bibinfo{pages}{237} (\bibinfo{year}{1986}).

\bibitem[{\citenamefont{Markidou et~al.}(2005)\citenamefont{Markidou, Shih, and
  Shih}}]{Markidou-2005-SME}
\bibinfo{author}{\bibfnamefont{A.}~\bibnamefont{Markidou}},
  \bibinfo{author}{\bibfnamefont{W.~Y.} \bibnamefont{Shih}}, \bibnamefont{and}
  \bibinfo{author}{\bibfnamefont{W.~H.} \bibnamefont{Shih}},
  \bibinfo{journal}{Rev. Sci. Instrum.}
  \textbf{\bibinfo{volume}{76}}, \bibinfo{pages}{064302}
  (\bibinfo{year}{2005}).

\bibitem[{\citenamefont{Kakavas}(1996)}]{Kakavas-1996-MPB}
\bibinfo{author}{\bibfnamefont{P.~A.} \bibnamefont{Kakavas}},
  \bibinfo{journal}{J. Appl. Polym. Sci. }
  \textbf{\bibinfo{volume}{59}}, \bibinfo{pages}{251} (\bibinfo{year}{1996}).

\bibitem[{\citenamefont{Rinde}(1970)}]{Rinde-1970-PRR}
\bibinfo{author}{\bibfnamefont{J.~A.} \bibnamefont{Rinde}},
  \bibinfo{journal}{J. Appl. Polym. Sci. }
  \textbf{\bibinfo{volume}{14}}, \bibinfo{pages}{1913} (\bibinfo{year}{1970}).

\bibitem[{\citenamefont{Goldstein and Salganik}(1974)}]{Goldstein-1974-BFS}
\bibinfo{author}{\bibfnamefont{R.~V.} \bibnamefont{Goldstein}}
  \bibnamefont{and} \bibinfo{author}{\bibfnamefont{R.~L.}
  \bibnamefont{Salganik}}, \bibinfo{journal}{Int. J. Fract.}
  \textbf{\bibinfo{volume}{10}}, \bibinfo{pages}{507} (\bibinfo{year}{1974}).

\bibitem[{\citenamefont{Audoly et~al.}(2005)\citenamefont{Audoly, Reis, and
  Roman}}]{Audoly-2005-CTS}
\bibinfo{author}{\bibfnamefont{B.}~\bibnamefont{Audoly}},
  \bibinfo{author}{\bibfnamefont{P.~M.} \bibnamefont{Reis}}, \bibnamefont{and}
  \bibinfo{author}{\bibfnamefont{B.}~\bibnamefont{Roman}},
  \bibinfo{journal}{Phys. Rev. Lett.} \textbf{\bibinfo{volume}{95}},
  \bibinfo{pages}{025502} (\bibinfo{year}{2005}).

\bibitem[{\citenamefont{Olson and Pollard}(1989)}]{Olson-1989-IPN}
\bibinfo{author}{\bibfnamefont{J.}~\bibnamefont{Olson}} \bibnamefont{and}
  \bibinfo{author}{\bibfnamefont{D.~D.} \bibnamefont{Pollard}},
  \bibinfo{journal}{Geology} \textbf{\bibinfo{volume}{17}},
  \bibinfo{pages}{345} (\bibinfo{year}{1989}).

\end{thebibliography}

\end{document}